\documentclass{aa}
\def\ltsim{\mathrel{<\kern-1.0em\lower0.9ex\hbox{$\sim$}}}
\def\gtsim{\mathrel{>\kern-1.0em\lower0.9ex\hbox{$\sim$}}}

\begin{document}

\title{The B3--VLA CSS sample. II: VLBA images at 18 cm}

\author{ D. Dallacasa \inst{1,2} \and S. Tinti \inst{1,2}\thanks{\emph{
Present address: }S.Tinti, SISSA, Via Beirut 2--4, I--34014 Trieste, Italy}\and C. Fanti 
\inst{3,2}  \and R. Fanti \inst{3,2} \and L. Gregorini \inst{3,2} \and 
C. Stanghellini \inst{4} \and M. Vigotti \inst{2} }

\institute {Dipartimento di Astronomia, Universit\'a di Bologna,
Via Ranzani 1, I--40127 Bologna, Italy \and Istituto di Radioastronomia del
CNR, via Gobetti 101, I--40129 Bologna, Italy \and Dipartimento di Fisica,
Universit\'a di Bologna, Via Irnerio 46, I--40126 Bologna, Italy \and Istituto 
di Radioastronomia del CNR, CP 141, I--96017 Noto SR, Italy} 

\offprints{D. Dallacasa}
\mail{ddallaca@ira.cnr.it}

\date{Received {{22 January 2002}  } / Accepted {{11 April 2002} } }

      %
      %
      \markboth{B3--VLA CSSs: VLBA}{Dallacasa et al.}
      \titlerunning{B3--VLA CSSs: VLBA}
      \authorrunning{Dallacasa  et al.}

\abstract{
VLBA observations at 18 cm are presented for 28 Compact Steep--spectrum 
radio Sources (CSSs) from the B3--VLA CSS sample.
These sources were unresolved in previous VLA observations at high 
frequencies or their brightness distribution was dominated by an unresolved 
steep spectrum component. More than half of them also showed a low frequency 
turnover in their radio spectrum.
The VLBA images display in most cases a compact symmetric structure. Only in a
minority of cases complex morphologies are present. 
\keywords{galaxies: active -- radio continuum:galaxies -- quasar: general }}

\maketitle

\section {Introduction}
The population of CSSs ({\bf C}ompact {\bf S}teep--spectrum {\bf S}ources) \&
GPSs ({\bf G}Hz {\bf P}eaked--spectrum {\bf S}ources) (see O'Dea, \cite{ode2}
for a review)
has received recently an increasing attention as at least the two sided
(or symmetric) objects of this class are considered to be the young stage
in the radio source evolution (Fanti et al. \cite{fan1}; Readhead et
al. \cite{read}), from the Compact (CSO, size $< 1~ h^{-1}$
kpc)\footnote{$H_0=100\,h$ km\,s$^{-1}$Mpc$^{-1}$,
$q_0=0.5$}  through
the Medium (MSO, 1 $h^{-1}$ kpc $<$ size $<$ 20 $h^{-1}$ kpc) to the
Large Symmetric Object (LSO, size $>$ 20 $h^{-1}$ kpc). A strong
support for this interpretation comes from the large expansion
velocities recently measured in a handful of sub-kpc sources (see
compilation in Fanti \cite{fan2} and references therein) and from the
determination of the radiative ages of a larger number of CSSs
spanning a range of linear size from sub--kpc to $\sim 20~h^{-1}$kpc
(Murgia et al. \cite{mur}). 

Starting from the seminal paper by Baldwin \cite{bald}, models for radio 
source evolution have been presented by several authors
which link the CSS/GPS population to the larger size sources,  (see e.g. 
Begelman \cite{beg}; Kaiser \& Alexander \cite{kai}).

All models essentially predict a decrease of the source luminosity with
time, and therefore with increasing source linear size.
A key test for the models is the ability to reproduce the distribution of
sources in the ``Radio Power -- Linear Size" ($P_{\rm R} - LS$) plane
(Baldwin \cite {bald}).
CSSs/GPSs seem to be in
the right proportion as compared with model expectations (Fanti et al.
\cite{fan1};
Readhead et al. \cite{read}). However, from the samples studied so far 
the  statistics  have been not very large and
suspicions were raised on some  ``irregularities" in the $LS$ 
distribution (O'Dea \& Baum, \cite{ode1}) and on the, perhaps too high, 
number of very small size CSSs, the GPSs (Polatidis et al.  \cite {pola};
Fanti \& Fanti \cite {elba}).

Recently we have selected a new sample of CSSs from the B3--VLA catalogue
(Vigotti et al.  \cite{vig1})
aimed at increasing significantly the statistics on sources with $LS$ in
the range $0.4~h^{-1} \leq LS$(kpc)$ \leq 20~h^{-1}$ (Fanti et
al. \cite{fan3}, hereafter referred to as Paper I). This sample,
consisting of 87 CSSs ($LS \leq 20$ $h^{-1}$ kpc), has VLA
observations at 1.4 GHz 
(A and C configurations),  5 and 8.4 GHz (both A configuration).
About 50 \% of these sources were not resolved or were poorly resolved
even at 8.4 GHz  
(beam size $\sim 0.2\arcsec$). For them two VLBI observing 
projects were undertaken with the VLBA and EVN \& MERLIN.

This paper reports on the results of the VLBA observations of the
most compact sources. The EVN \& MERLIN observations instead are
presented in a companion paper by Dallacasa et al. (\cite{dalla},
Paper III). 

\begin{table*}[htbp]
\caption{The VLBA Sample: See text for a description of the columns.}
\begin{center}
\begin{tabular}{lcclrrrrrrl}
\hline
\hline
Source   & Id &$ m_R$ &   z~~~ &  $LAS$  & Log P$_{0.4 \rm GHz}$&
~S$_{1.67}^{\rm VLA}$&~~S$_{1.67}^{\rm VLBA}$&$LAS_{\rm VLBA}$& $LLS~~$&Morph. \\
         &    &   &       &~mas &W/Hz $h^{-2}$ &mJy&mJy&mas~~~~&kpc $h^{-1}$&        \\
~~~~(1)  & (2)&(3)& ~~(4)~& (5)~   &    (6)  &(7)~~&(8)~~&(9)~~~~~& (10)~~& (11) \\
\hline
\hline
0039+373\ \       &  G &      & 1.006  &100 &   27.37&  784&  779 & 120~~~~ &      0.50 &CSO \\
0147+400\ \       &  E &      &           &100 &$>$26.6 &  631&  563 & 95~~~~ &$\sim$0.40 &scJ?$^\ddag$ \\
0703+468\ \       &  E &      &           &80 &$>$26.6 & 1392& 1356 &  75~~~~ &$\sim$0.32 &CSO \\
0800+472 \ $\ast$ &  E &      &           &1000 &$>$26.8 &  769&  447 & 160~~~~ &$\sim$0.67 &scJ?$^\ddag$\\ 
0809+404 \ $\ast$ &  G &      & 0.551  &1200 &   26.90&  929&  522 &
80~~~~ &      0.28 & ~~?\\ 
0822+394\ \       &  G &      & 1.180  &50 &   27.33& 1003&  942 &  70~~~~ &      0.30 & CSO \\ 
0840+424A\        &  E &      &           &80 &$>$26.8 & 1243& 1187 & 135~~~~ &$\sim$0.57 &CSO \\ 
1007+422 \ $\ast$ &  E &      &           &130 &$>$26.4 &  370&  313 & 265~~~~ &$\sim$1.12 &MSO \\ 
1008+423\ \       &  E &      &           &50 &$>$26.5 &  521&  498 & 115~~~~ &$\sim$0.48 &CSO\\ 
1016+443\ \       &  G & 19.7 & 0.33~  R&110 &   26.11&  287&  266 & 155~~~~ &      0.44 &CSO \\ 
1044+454 \ $\ast$ &  G & 24.8 &  4.10~  K&1000 &   28.82&  352&  311 &
175~~~~ &      0.55 &~~? \\ 
1049+384 \ $\ast$ &  G & 20.9 & 1.018  &100 &   27.14&  574&  502 & 210~~~~ &      0.89 &CSO?\\ 
1133+432\ \       &  E &      &           &70 &$>$26.3 & 1247& 1265 &  45~~~~~&$\sim$0.20 &CSO \\ 
1136+383 \ $\ast$ &  E &      &           &50 &$>$26.4 &  400&  377 &  70~~~~ &$\sim$0.28 &CSO \\ 
1136+420\ \       &  G & 21.7 & 0.829  &1000 &$^a~26.9$& 402&  279 &
120~~~~ &      0.49 &~~? \\ 
1159+395\ \       &  G & 23.4 & 2.370  &50 &   27.57&  535&  498 &  70~~~~ &      0.26 &CSO \\ 
1225+442\ \       &  G & 18.2 &  0.22~  R&200 &   27.50&  316&  291 & 420~~~~ &      0.95 &CSO \\ 
1242+410\ \       &  Q & 19.7 & 0.811  &40 &   27.08& 1215& 1126 &  70~~~~ &      0.29 &CSO?\\ 
1314+453A$\ast$   &  G & 21.8 & 1.544  &160 &   27.77&  574&  493 & 185~~~~ &      0.79 &CSO \\ 
1340+439\ \       &  E &      &           &70 &$>$26.5 &  447&  397 & 130~~~~ &$\sim$0.56 &CSO \\ 
1343+386 \ $\ast$ &  Q & 17.5 & 1.844  &110 &   27.70&  793&  717 & 130~~~~ &      0.54 &CSO \\ 
1432+428B$\ast$   &  E &      &           &40 &$>$26.3 &  809&  706 &  50~~~~ &$\sim$0.21 &CSO \\ 
1441+409\ \       &  E &      &           &100 &$>$26.7 &  834&  768 & 130~~~~ &$\sim$0.56 &CSO \\ 
1449+421\ \       &  E &      &           &80 &$>$26.9 &  639&  592 & 120~~~~ &$\sim$0.50 &CSO \\ 
2304+377\ \       &  G & 19.7 &  0.40~  R&100 &   26.44& 1299& 1217 & 150~~~~ &      0.49 &CSO \\ 
2330+402\ \       &  E &      &           &70 &$>$26.7 &  730&  705 & 100~~~~ &$\sim$0.43 &CSO \\ 
2348+450 \ $\ast$ &  G &      & 0.978 &200 &   27.36&  630&  520 & 310~~~~ &      1.32 &MSO \\ 
2358+406\ \       &  E &      &           &80 &$>$26.8 & 1159& 1161 & 120~~~~ &$\sim$0.47 &CSO \\  
\hline
\hline
\end{tabular}
\end{center}
\medskip
$^a$ the value given in Fanti et al. 2001 is wrong

$^\ddag$ steep-core Jet: elongated structure, very asymmetric in brightness, where the
brightest component is at one source extremity and has a steep spectrum.
\label{tabsample}
\end{table*}

\begin{figure}
\vspace{9 truecm}
\includegraphics{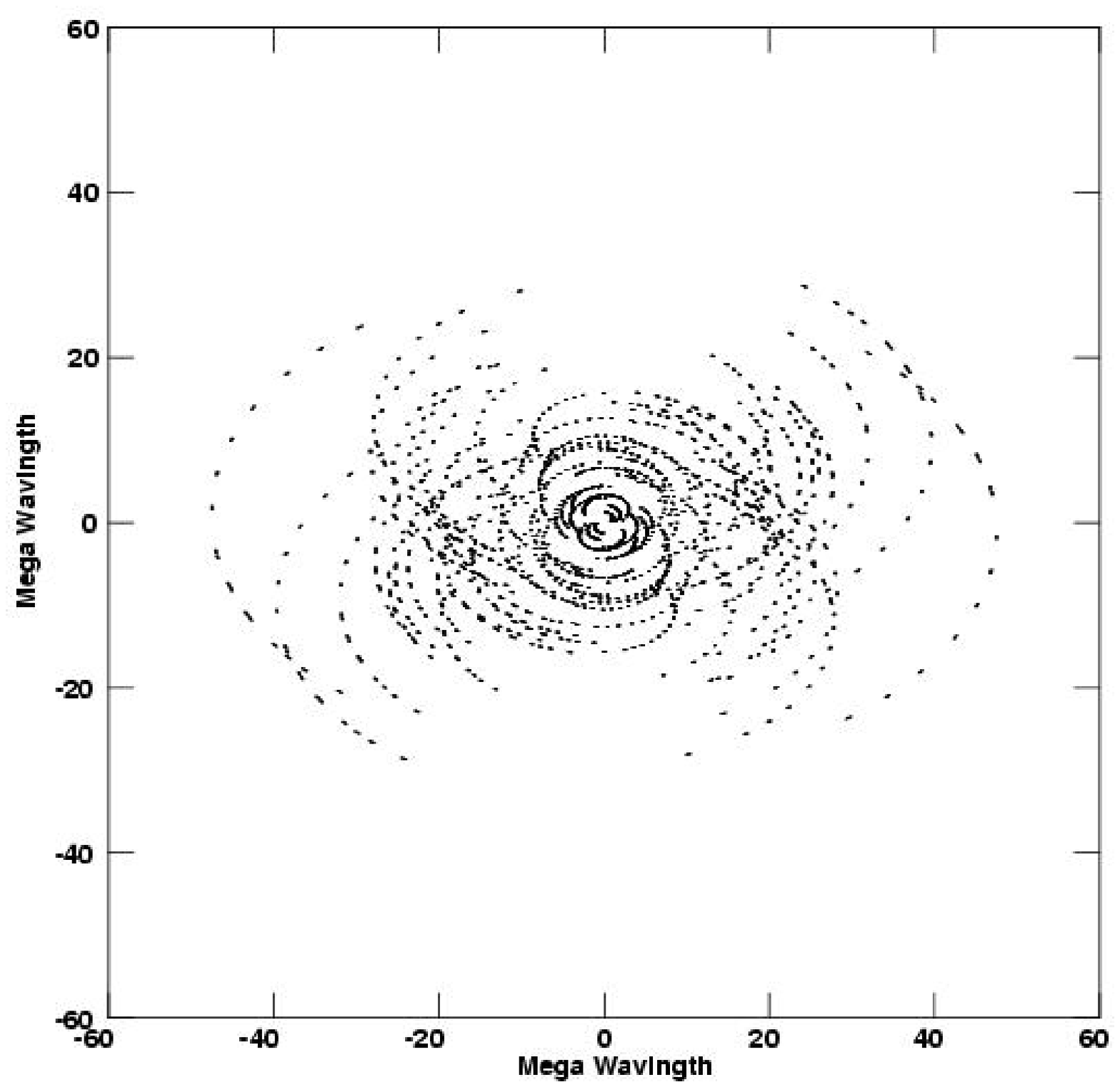}
\caption{Typical $uv$ coverage}
\label{uvcov}
\end{figure}

\section{The Source list and VLBA observations}
Among the twenty--eight sources selected for VLBA 
observations, twenty-four are unresolved or slightly resolved
(angular size $\leq 0.2\arcsec$) in the VLA 8.4 GHz observations. 
The remaining four sources  are dominated in flux density by an
unresolved or slightly resolved component, with the additional
presence of some weaker emission (bringing the overall size up to 
$\sim 1\arcsec$, Table
\ref{tabsample}) whose connection with the compact feature is unclear.
In addition about half of the sample sources present a low frequency turnover 
in the total spectrum which is usually considered an indication
of compactness in some sub--component. All these sources, therefore,
deserved further investigation.

The selected sample  is presented in Table \ref{tabsample}.
The content of the table is the following.

\smallskip
\noindent
{\it Column 1} -  Source name; an ``$\ast$'' refers to a note (Sect.
\ref{comm});

\smallskip
\noindent
{\it Column 2} - Optical Identification (Id) from Paper I (G = galaxy, 
Q = quasar, E = no known optical counterpart);

\smallskip
\noindent
{\it Column 3} - R magnitude;

\smallskip
\noindent
{\it Column 4 } - redshift; ``K'' and ``R'' indicate that the
redshift is estimated by photometric measurements in the
respective optical band (see Paper I for details);

\smallskip
\noindent
{\it Column 5} - VLA Largest Angular Size ($LAS$) in mas, from Paper I;

\smallskip
\noindent
{\it Column 6} - log $P_{\rm 0.4 GHz}$ ($P$ in W/Hz $h^{-2}$); for E
sources lower limits to the radio power have been computed
assuming  $z = 0.5$ (see also Paper~I for a wider discussion); 

\smallskip
\noindent
{\it Column 7} - flux density (mJy), interpolated at 1.67~GHz 
from the 1.4~GHz NVSS (Condon et al. \cite{nvss}) and the 5~GHz VLA
data from Paper I; 

\smallskip
\noindent
{\it Column 8} - VLBA 1.67 GHz flux density (mJy), measured on the images 
presented here (see Sect.  3);

\smallskip
\noindent
{\it Column 9} - VLBA Largest Angular Size (mas);

\smallskip
\noindent
{\it Column 10} - Largest Linear Size ($LLS$, kpc$~h^{-1}$) from VLBA data;
for E sources $z = 1.05$ was assumed (see Paper I for details)
and $LLS$, given with a leading ``$\sim$'', must be considered a
lower limit;

\smallskip
\noindent
{\it Column 11} - overall morphology (see Sect. \ref{morph})

\smallskip
The VLBA observations were carried out in February 2000 at the
frequency of 1.67 GHz, with a recording band-width of 32 MHz at
64Mbps. Each source was observed for a total of about 1 hour, 
spread into 10 to 15 short scans in order to improve the $uv$
coverage. A typical example of $uv$ coverage is given in fig. \ref{uvcov}. As no single VLA antenna was available, the shortest baseline
is Pie Town--Los Alamos  ($\sim$1.3~M$\lambda$). This may have
caused some flux density losses in the more extended sources/components.

\begin{figure*}
\vspace{9 truecm}
\includegraphics{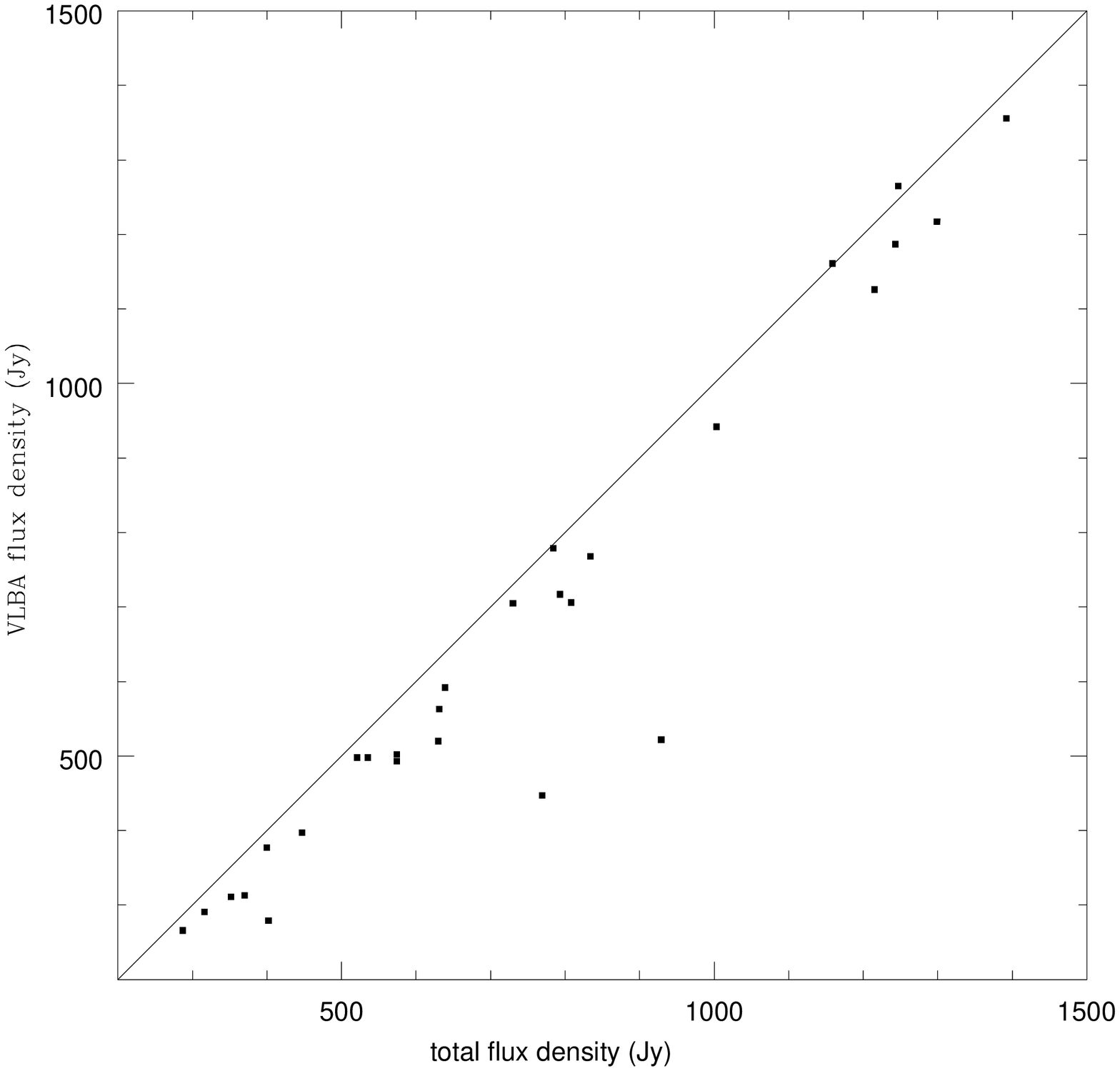}
\includegraphics{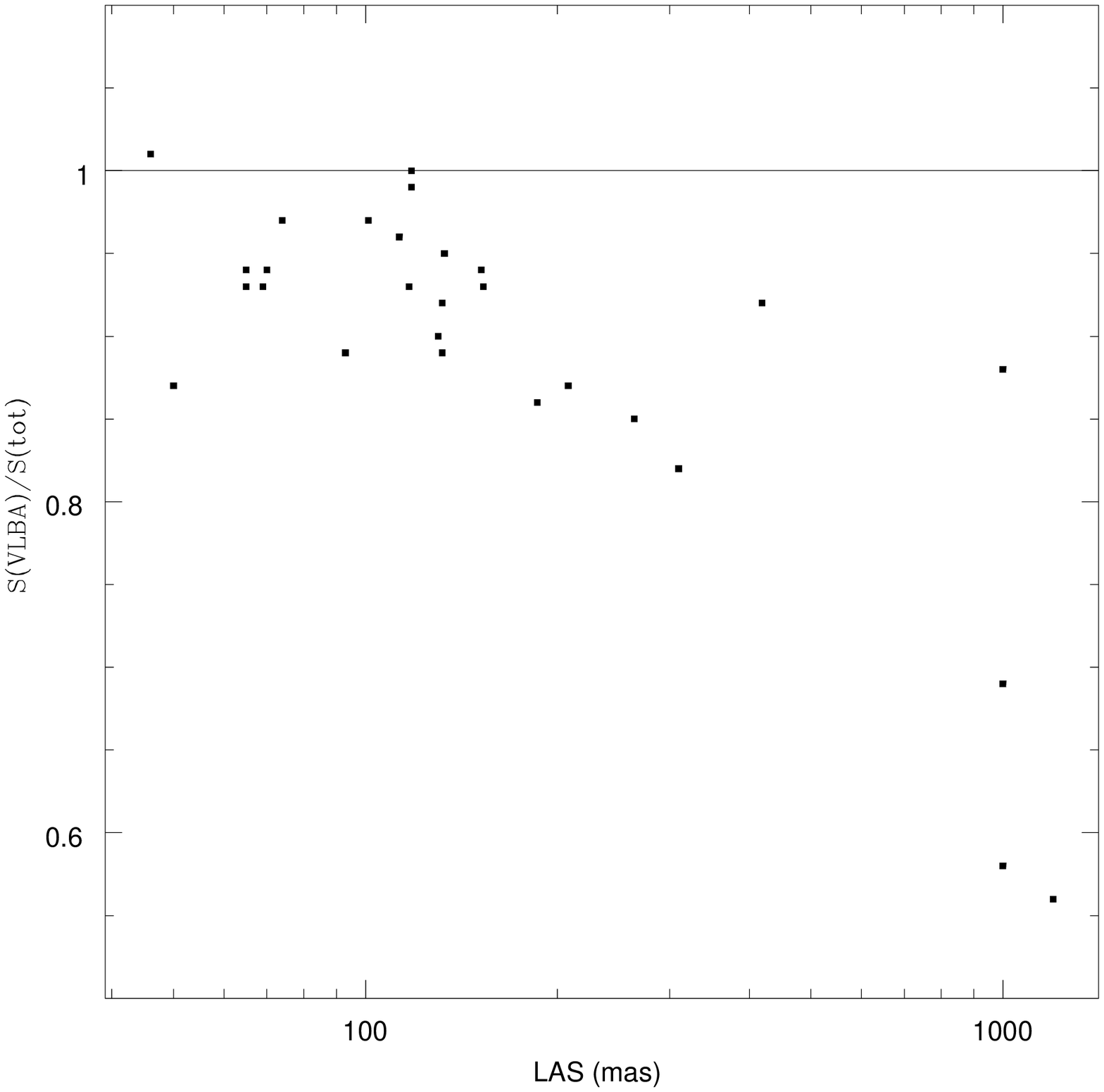}
\caption{{\it (left):} Comparison between the VLBA integrated flux
densities and total flux densities interpolated from the
literature. The line has a slope of one. {\it (right):} the ratio 
between the VLBA and the total flux density is shown as a function
of the overall source angular size} 
\label{fluxes}
\end{figure*}

\section{Data reduction}

The data were correlated with the NRAO processor  at Socorro.
Amplitude calibration was obtained using the system temperature 
measured at the various antennas
during the observations and the gain information provided
by NRAO. Consistency checks were performed with OQ208 and DA193, whose
amplitudes  were found to be generally within 2 \% of what expected.

Standard editing and fringe--fitting were applied. Most sources provided
fringes with high signal--to--noise ratio on all baselines. 

Radio images were obtained using AIPS, after a number of phase
self--calibrations, ended by a final amplitude self--calibration. The
last step was made with great care, as the gain self--calibration
tends to depress the total flux density when extended components are
poorly sampled in $uv$ coverage. Generally the gain corrections from
self--calibration were $<3$ \%. In those sources in which they
were $\geq 3$ \%, or in which the total flux density decreased by more
than 1 -- 2 \%,  they were not considered. 

The r.m.s. noise level (1 $\sigma$), measured on the images far from the
sources, is in the range of 0.1--0.2 mJy/beam, comparable to the
expected thermal noise. The dynamic range, defined as the ratio of
peak brightness to 1 $\sigma$, is $\sim$  700 on average, ranging from
300 to a few thousands. The typical resolution is $\sim 4 \times 8$
mas. We must remark that the $LAS$ determined on the VLBA images
(Table~1, Col. 9) does not take into account the eventual presence
of components completely resolved out by the VLBA observations. We
note also that it may exceed the VLA rough estimate due to the
source structure (e.g. a double edge-brightened source will appear
smaller when fitted by a single Gaussian with a resolution smaller
than the separation between components).

\begin{figure*}
\vspace{23 truecm}
\includegraphics{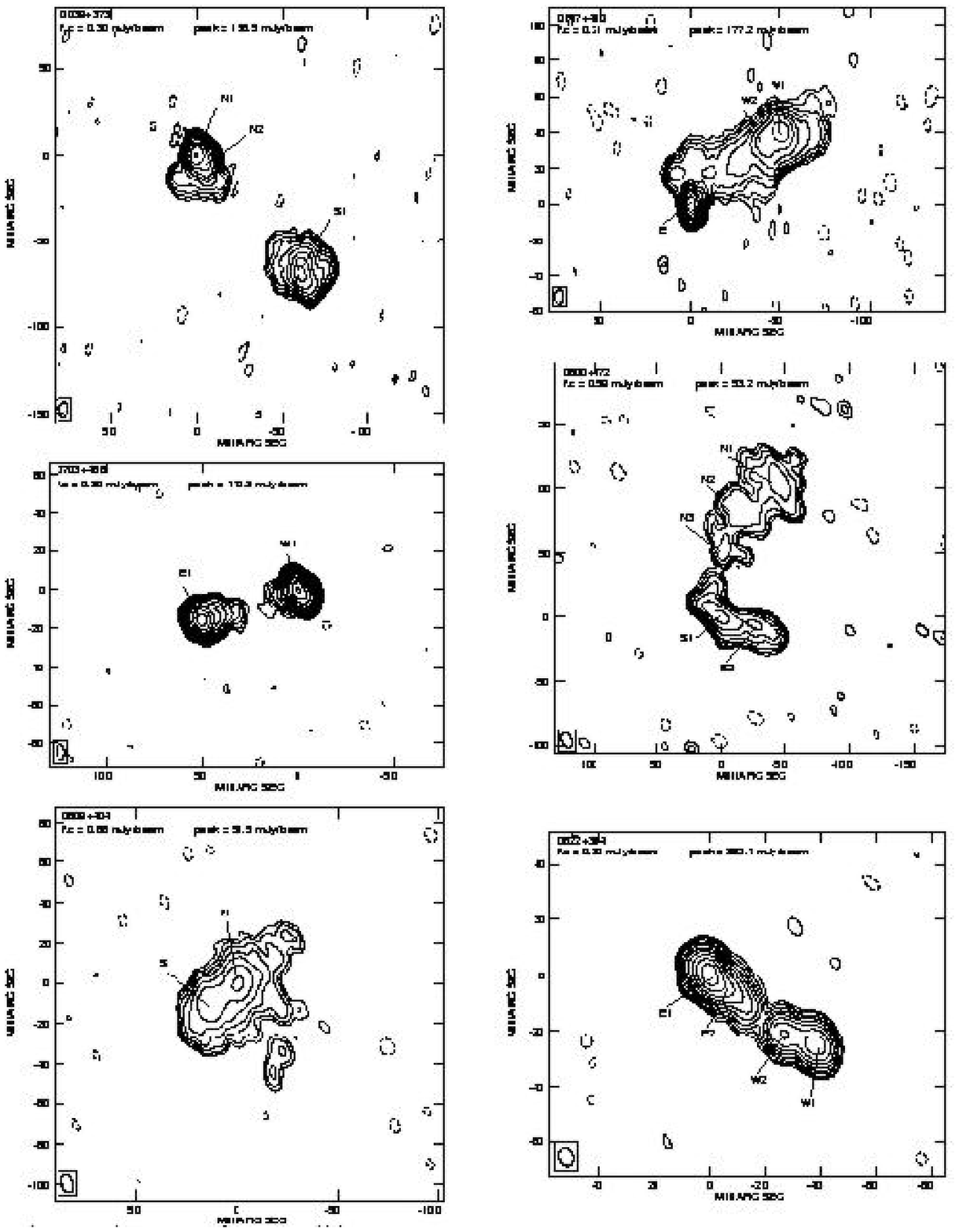}
\caption{VLBA images: the first contour (f.c) is  generally
three times the r.m.s. noise level on the image; contour levels
increase by a factor of 2; the restoring beam is shown in the bottom
left corner in each image.}
\label{images}
\end{figure*}

The total flux density  ``seen'' by VLBA (Table 1, Col.~8) has been
measured by integration over the source images using the AIPS task
TVSTAT. The brightest sub--structures of each source were fitted by a
gaussian model (task JMFIT) which allows to obtain total flux density,
the beam--deconvolved Half Maximum Widths (HMW) and major axis position angle (p.a.).
Sub--components are referred to as North ($N$), South ($S$),
East ($E$), West ($W$) and Central ($Ce$). When a component is split into more 
pieces, a digit (1,2, etc.) is added (e.g. $N1, N2$). Other symbols used are
self--explanatory. The source images are shown in Fig.~\ref{images}.
The labels are marked on the images, next to each sub--component.
The parameters of the brightest sub--components are given in Table 
\ref{comparam}, which has the following layout:

\smallskip
\noindent
{\it Columns 1, 2} --  Source name and sub--component label; an
``$\ast$'' indicates that the component is extended and its flux
density is obtained by TVSTAT (see text);

\smallskip
\noindent
{\it Column 3} --  VLBA flux density;

\smallskip
\noindent
{\it Column 4} --  deconvolved angular sizes of major and minor axis of a Gaussian
component and position angle of the major axis as estimated using
the AIPS Gaussian--fitting task JMFIT.

For more extended features, marked by an ``$\ast$'' in Table~2,  flux 
densities are evaluated by means of TVSTAT. For extended structures
underlying  more compact ones, 
the flux density is obtained as difference between the integrated flux 
density and the sum of the sub--component gaussian fit flux densities. 

We have checked the integrated flux densities in our images with the
low resolution flux densities obtained by interpolating to our
observing frequency the values from the NVSS (Condon et
al. \cite{nvss}) and the from VLA data at 5 GHz from Paper I. The
comparison is displayed in Fig. \ref{fluxes}.  

There is clearly a systematic flux density difference, typically
$\gtsim$ 5 \%, which could be partially ascribed to calibration
uncertainties. 
For the more extended objects there are larger flux density losses,
clearly due to the absence of short uv--spacings (see, as an
example, Fig. \ref{fluxbas}).
The two sources with the largest deviations in
Fig.~\ref{fluxes} (0800+472 and 0809+404) do possess indeed another  
component detected by the VLA but completely resolved out by the VLBA
baselines (see also next section).   

\setcounter{figure}{2}
\begin{figure*}
\vspace{23 truecm}
\includegraphics{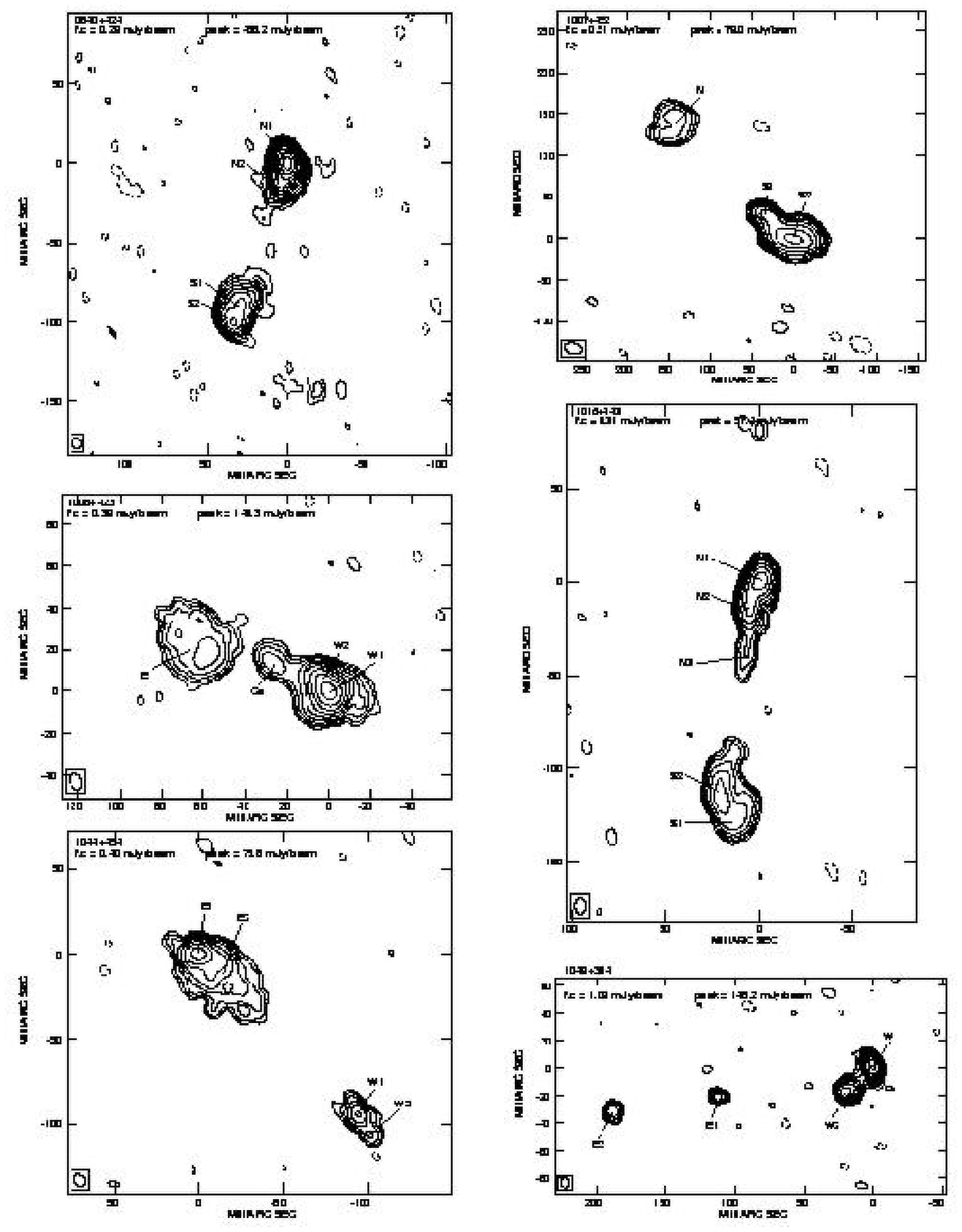}
\caption{VLBA Images (cont.)}
\end{figure*}

\setcounter{figure}{2}
\begin{figure*}
\vspace{23 truecm}
\includegraphics{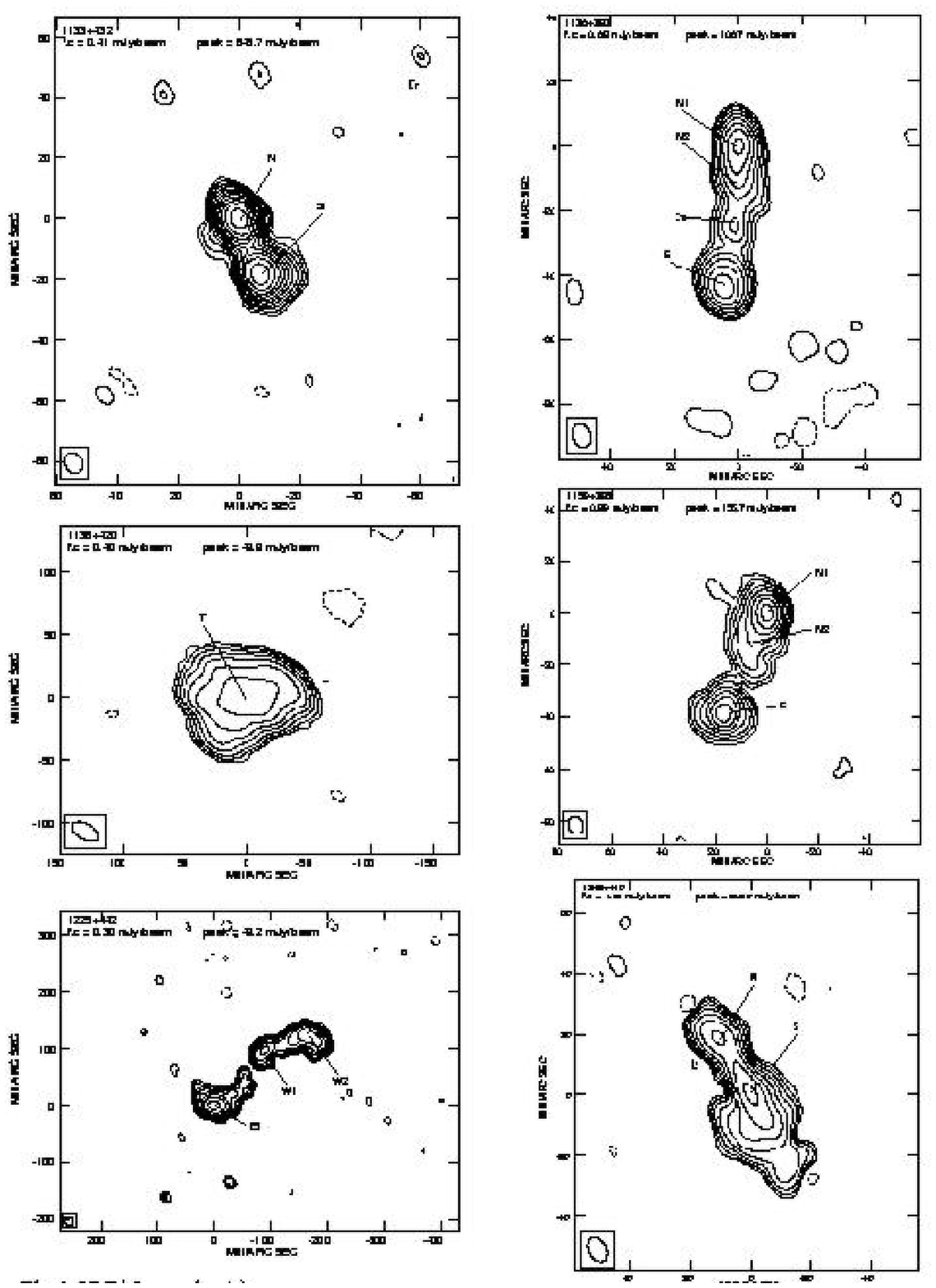}
\caption{VLBA Images (cont.)}
\end{figure*}

\setcounter{figure}{2}
\begin{figure*}
\vspace{23 truecm}
\includegraphics{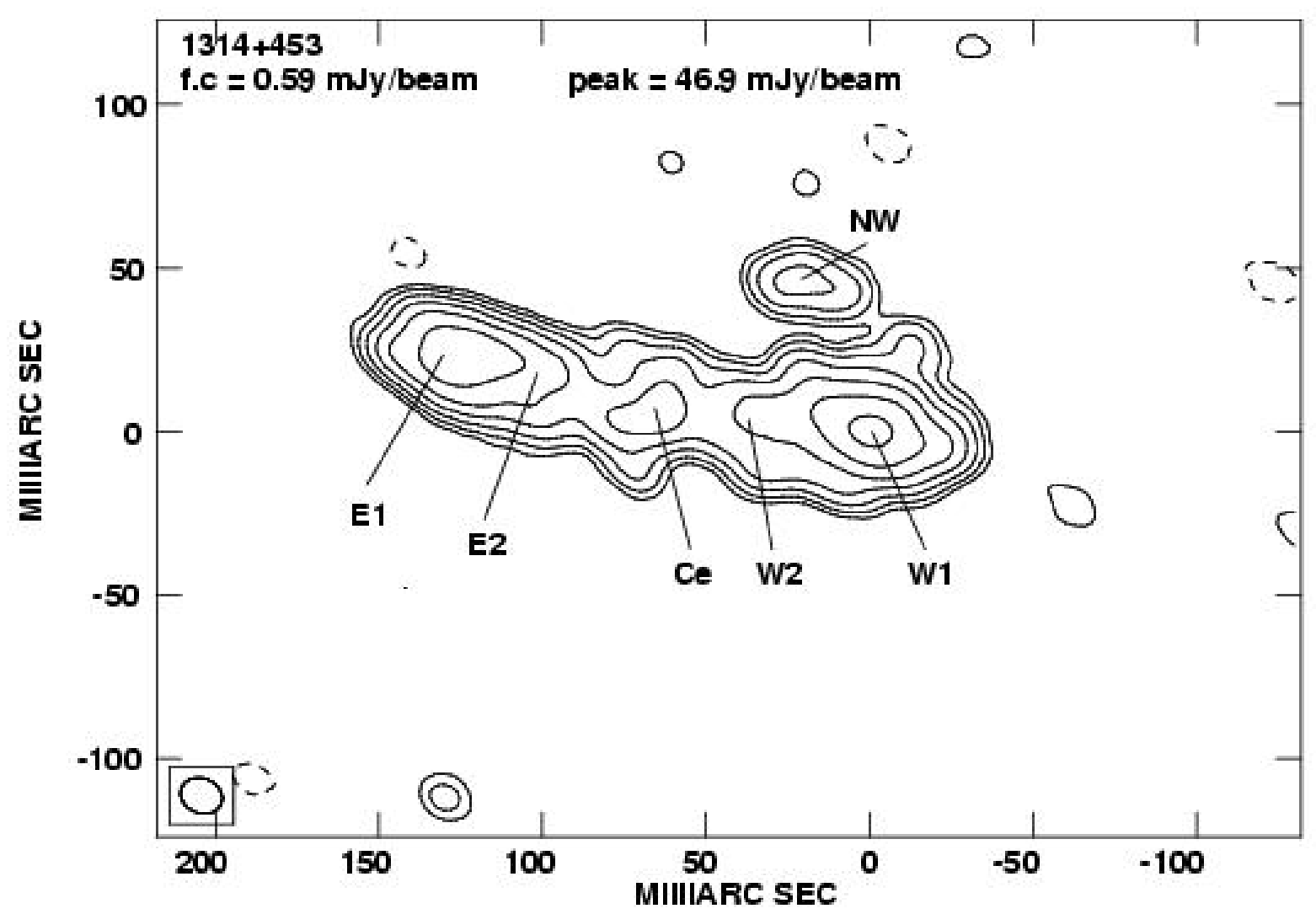}
\includegraphics{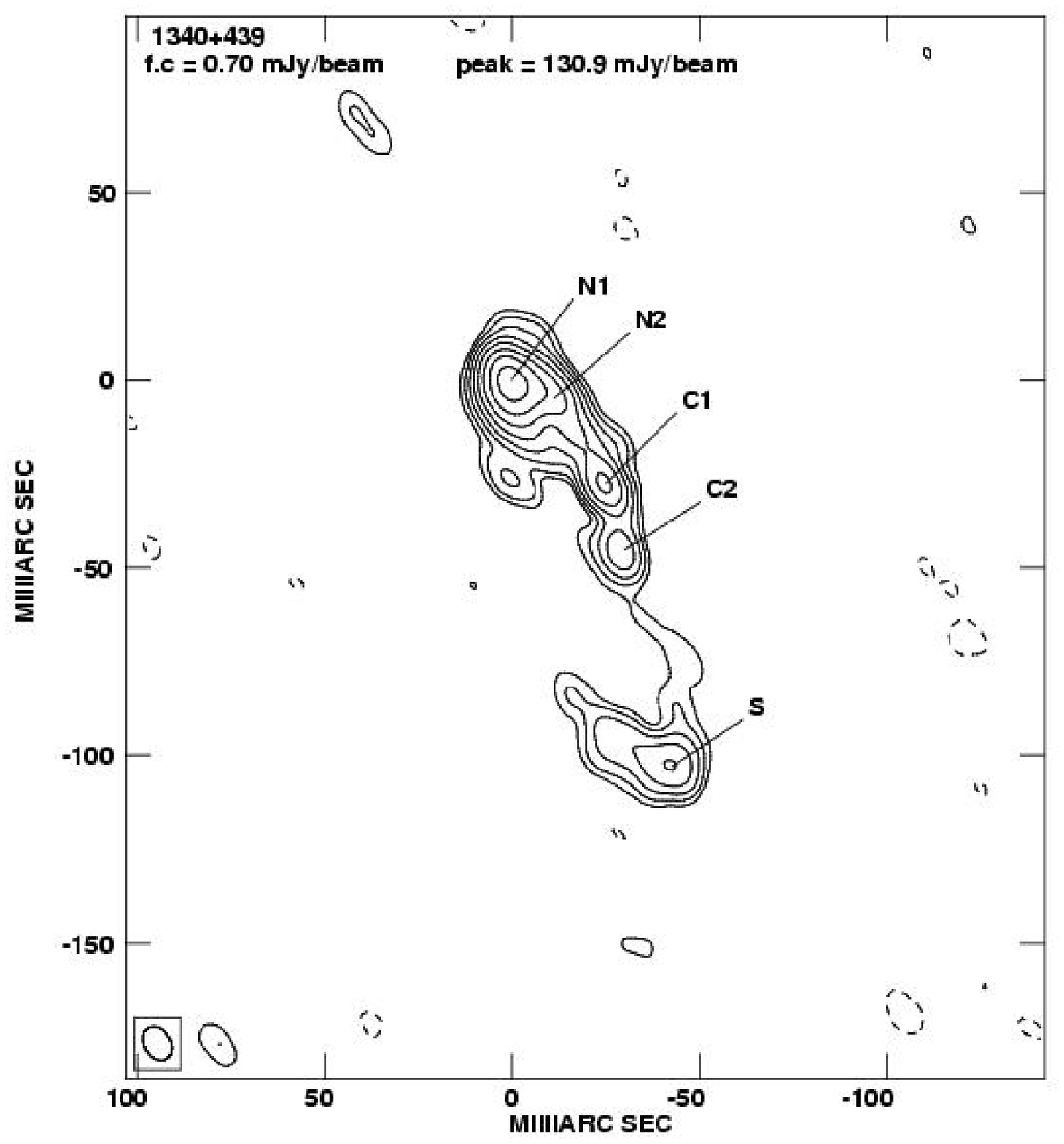}
\includegraphics{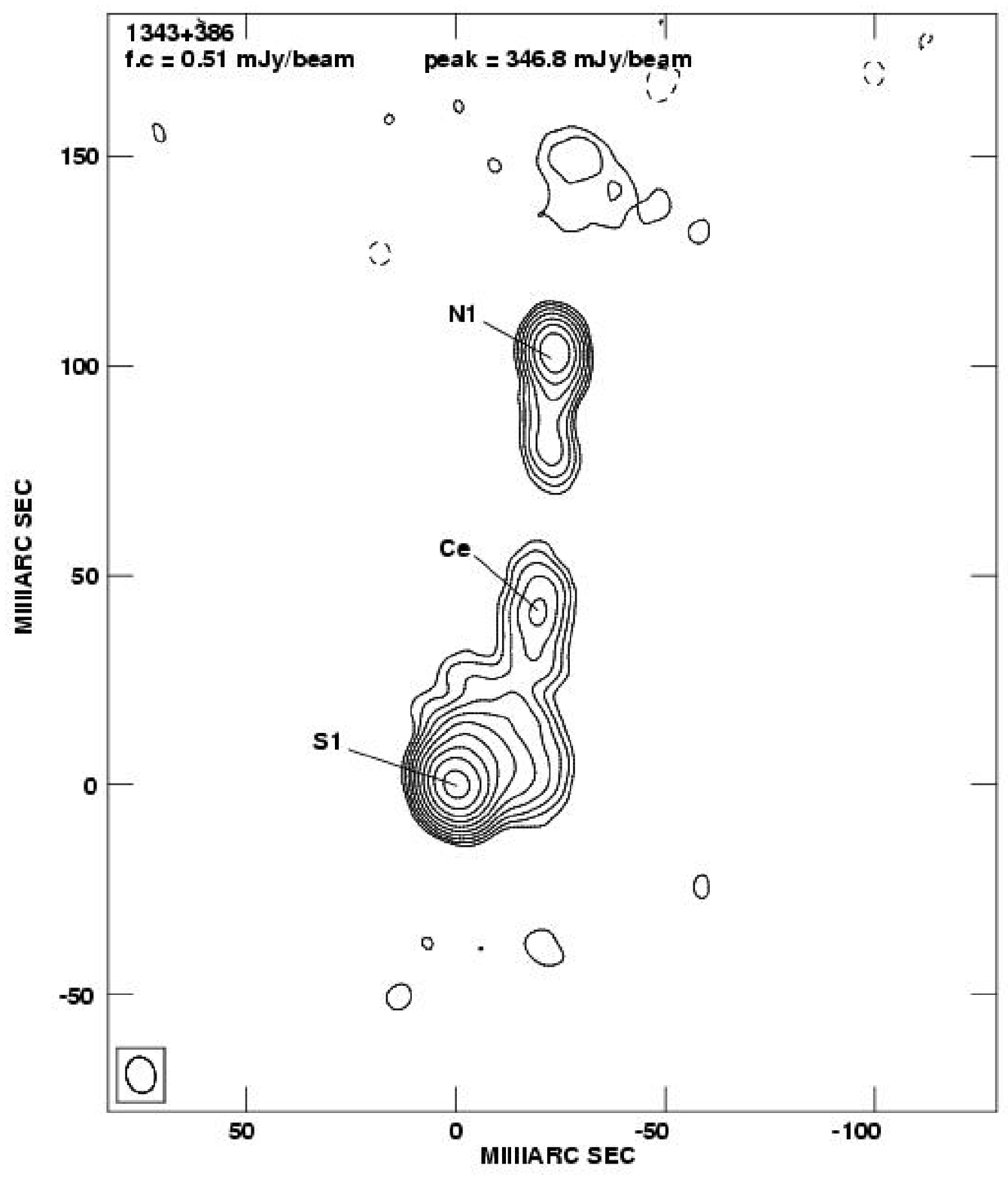}
\includegraphics{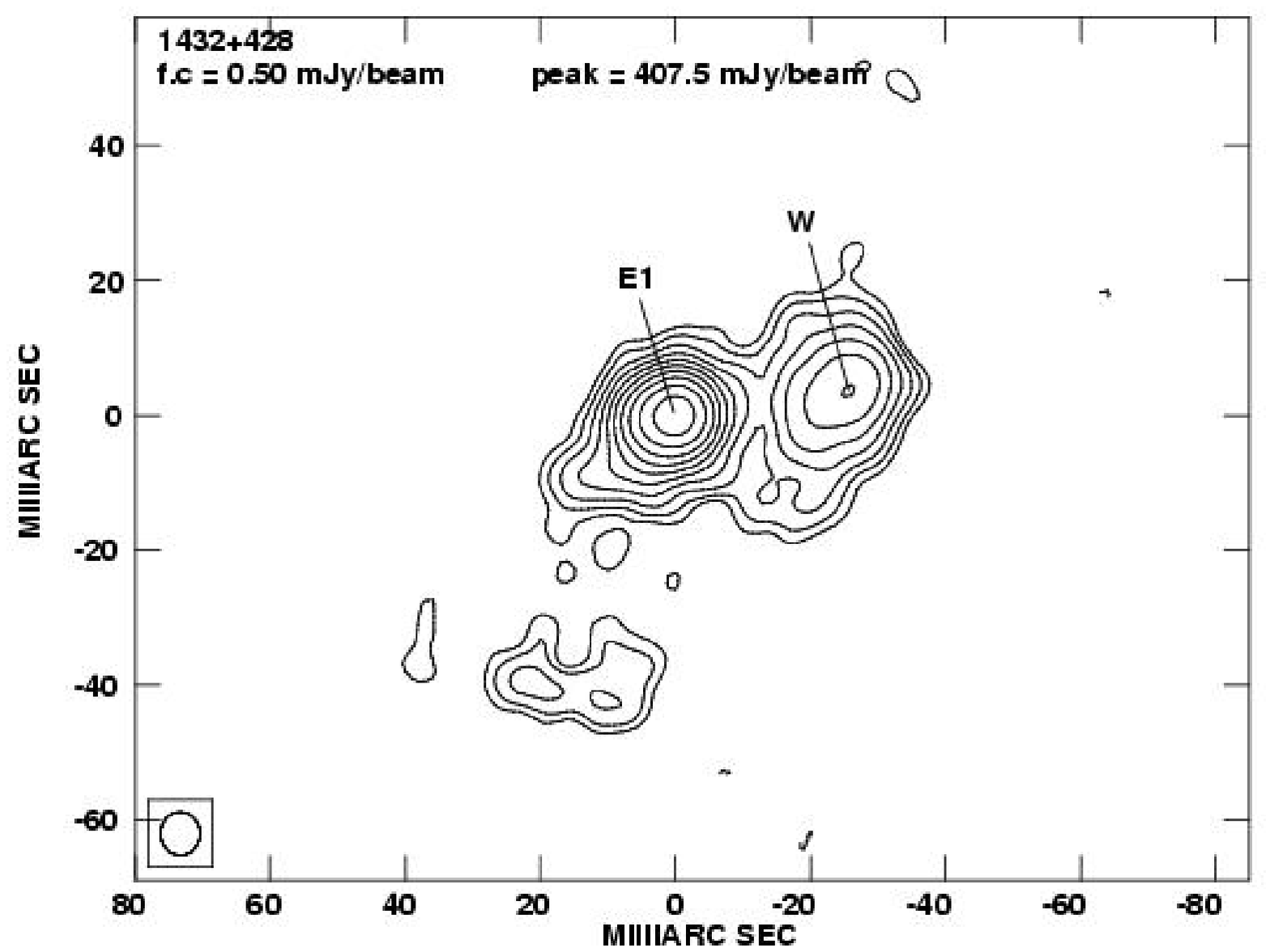}
\includegraphics{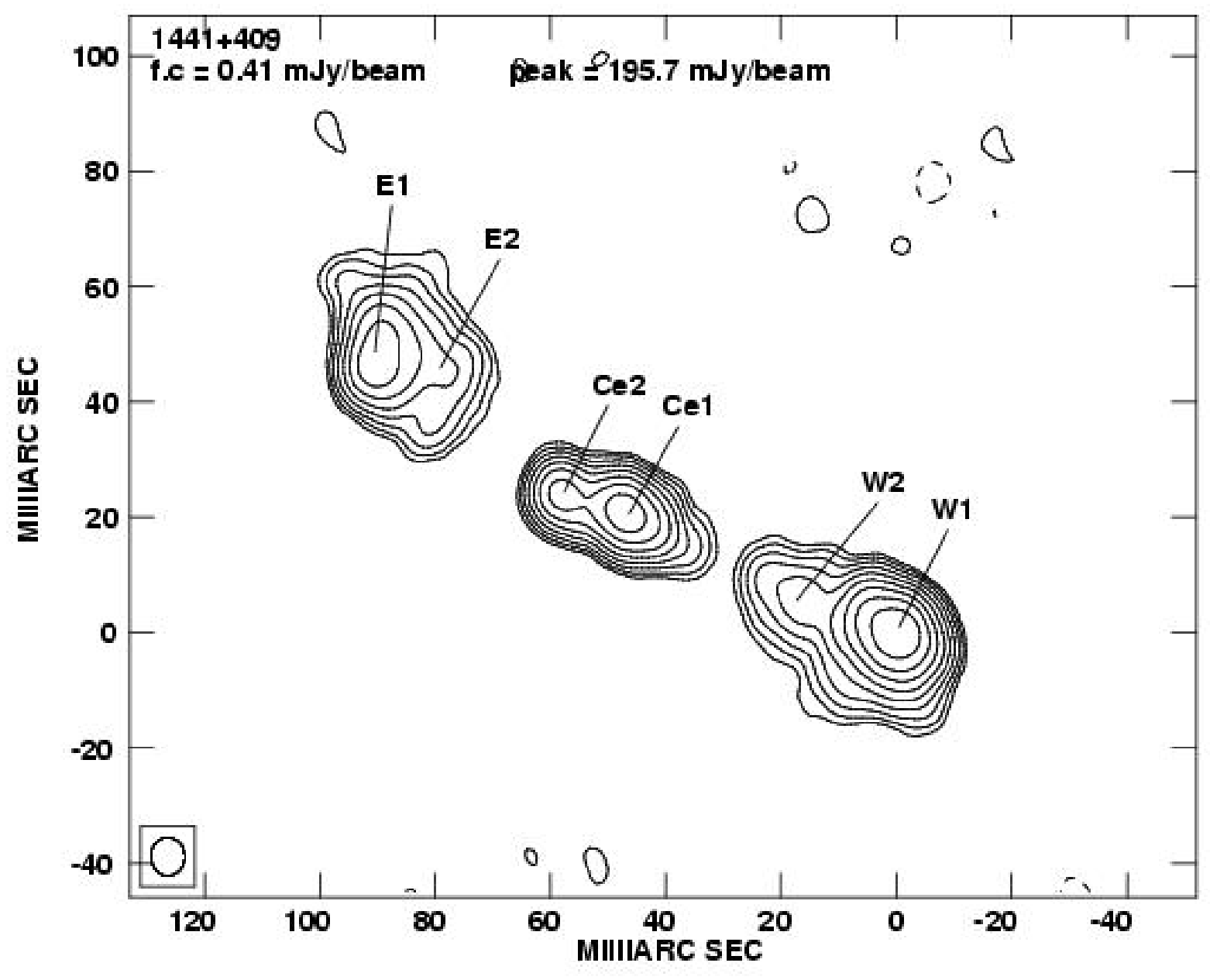}
\includegraphics{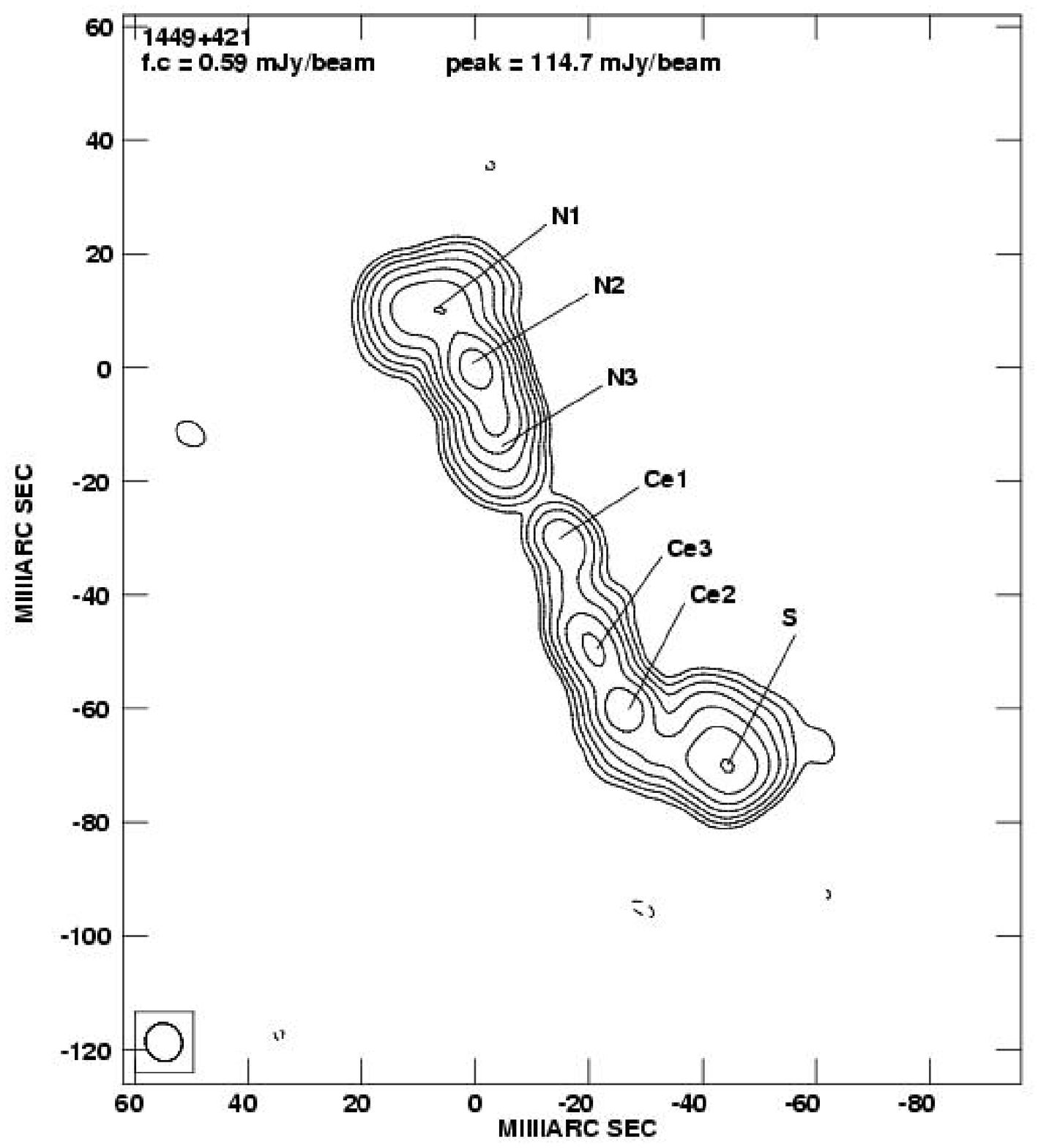}
\caption{VLBA Images (cont.)}
\end{figure*}

\setcounter{figure}{2}
\begin{figure*}
\vspace{16.5 truecm}
\includegraphics{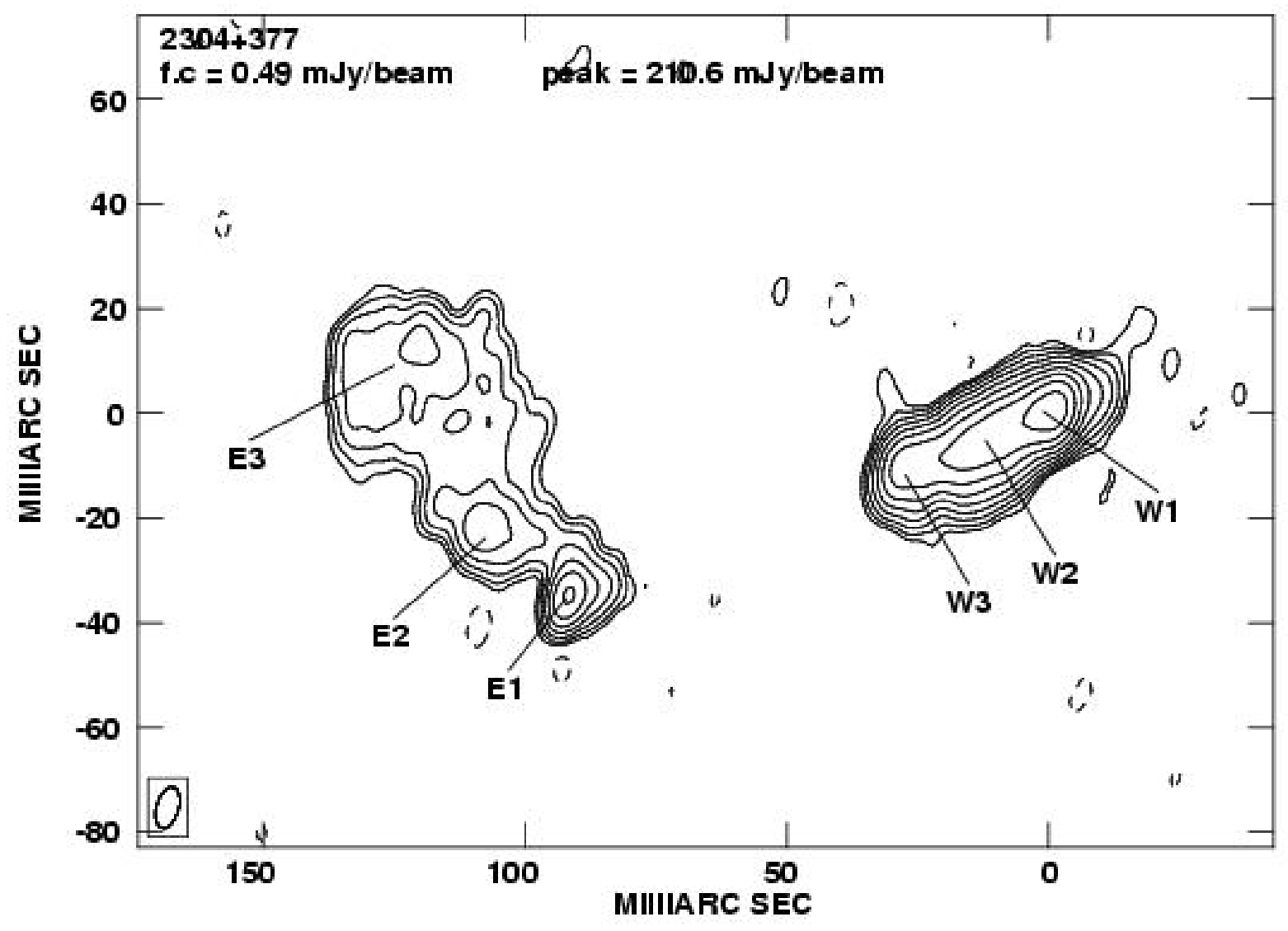}
\includegraphics{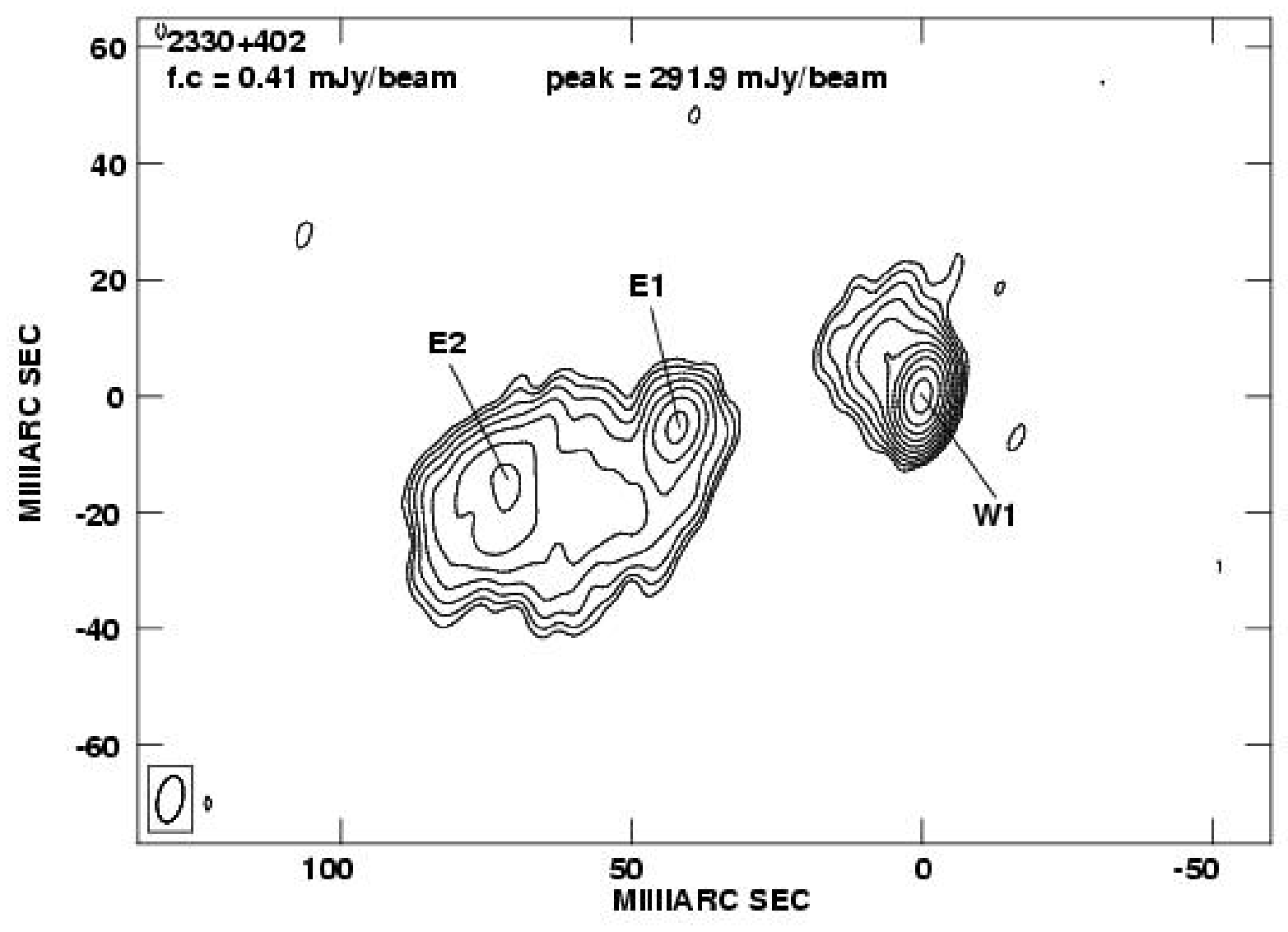}
\includegraphics{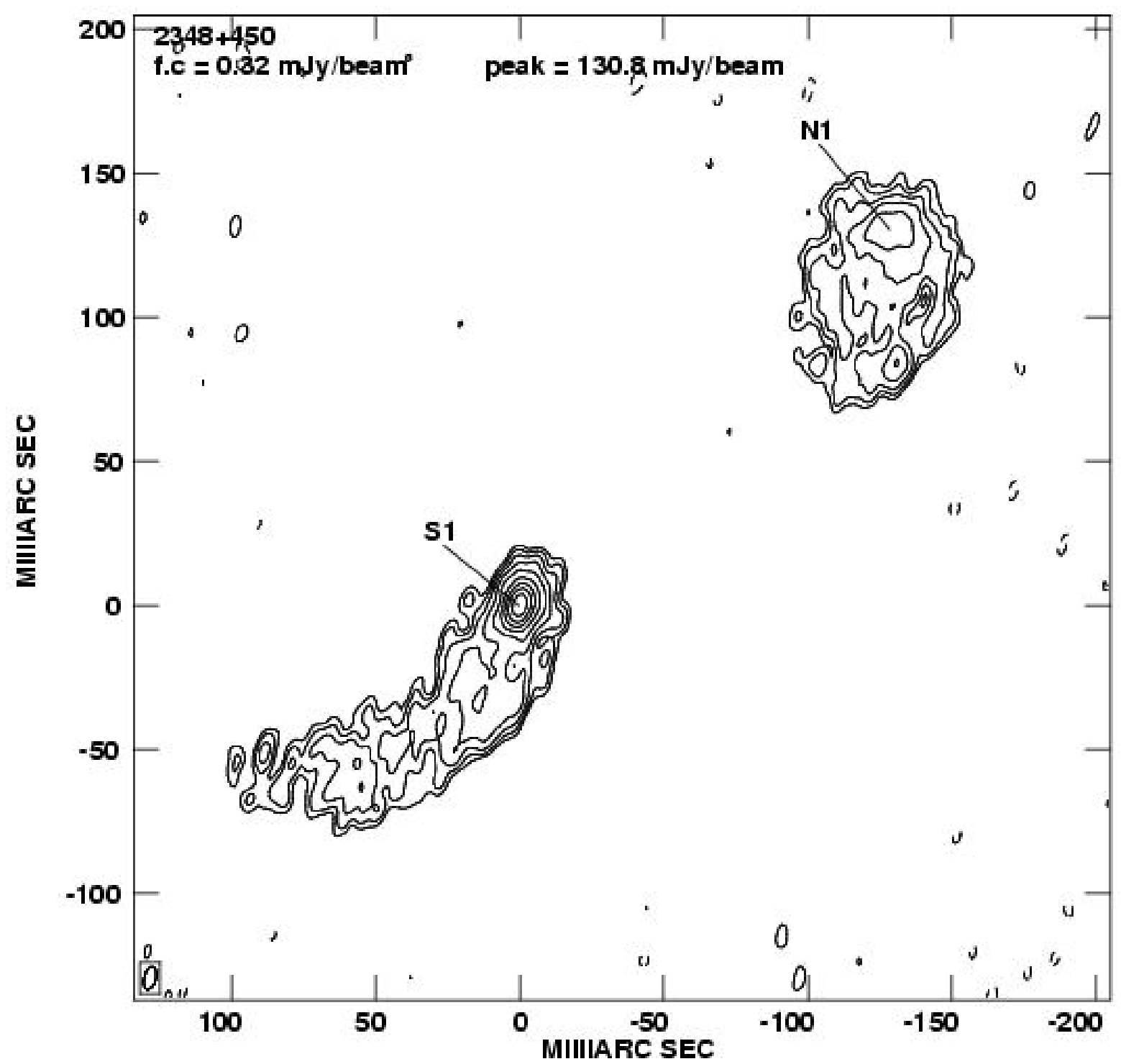}
\includegraphics{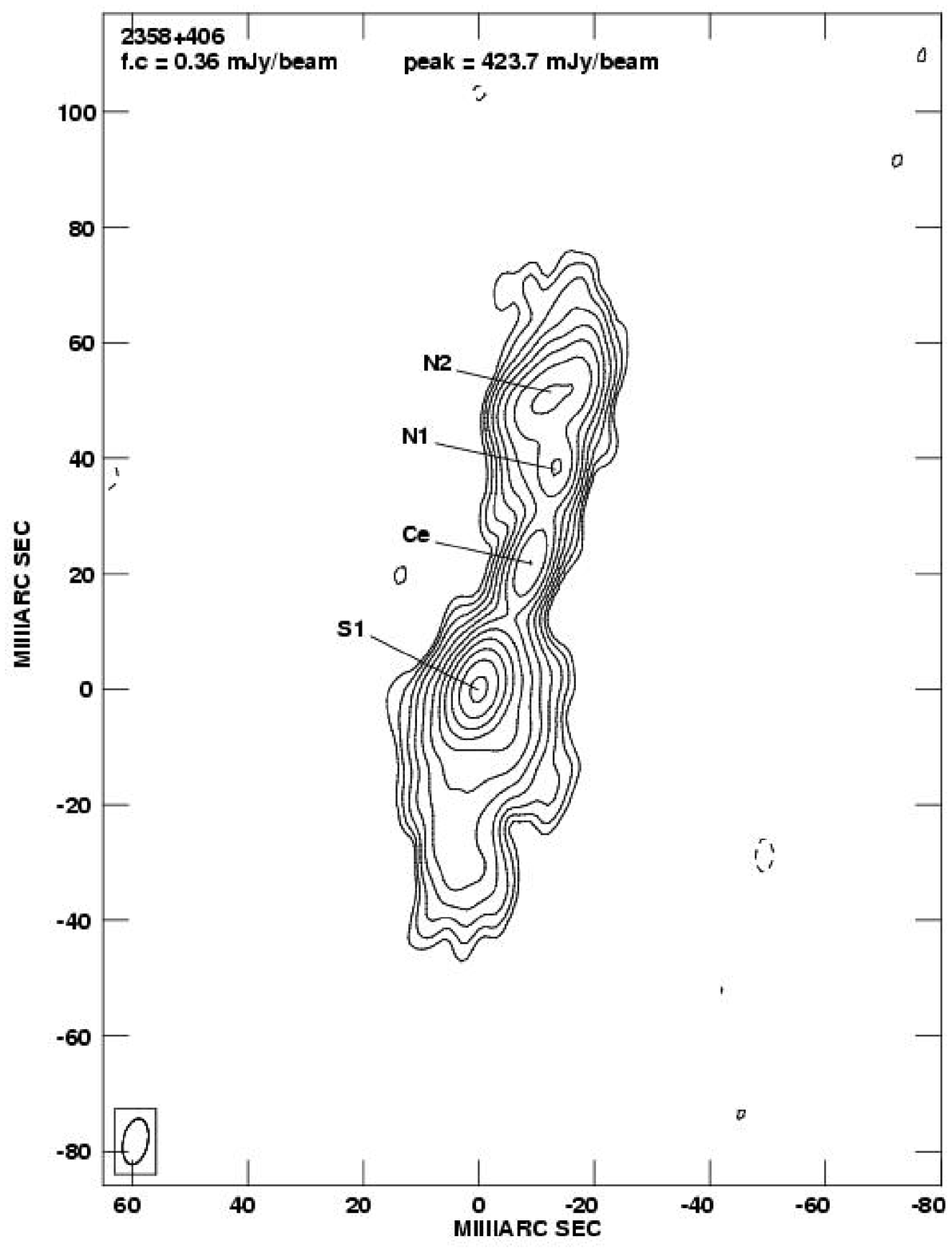}
\caption{VLBA Images (cont.)}
\end{figure*}

\begin{table*}
\begin{center}
\caption{Observational parameters of source components}
\scriptsize
\begin{tabular}{|ccrrrr|ccrrrr|}
\hline
\hline
Source & &  S &  $\theta_1$ & $\theta_2$ & (p.a) & Source & &  S &  $\theta_1$ & $\theta_2$ & (p.a) \\
         & &  (mJy) &  mas & mas & deg     &     & &  (mJy) &  mas & mas & deg \\ 
    (1)  & (2)    & (3) &  --- &  (4) & ---&     (1)  & (2)    & (3) &  --- &  (4) & --- \\
\hline
\hline
0039+373 & $ N1 $ & 229 &  4.7 &  4.3 & 101& 1225+442 & $ E1 $ & 138 &26.6 &  15.1 &  88 \\ 
         & $ N2 $ & 143 & 11.0 &  6.8 &  17&          &$E^\ast$ & 28 &     &       &     \\ 
         &$N^\ast$&  17 &      &      &    &          & $ W1 $ & 94 & 46.6 &  23.3 &  71 \\
	 & $ S1 $ & 303 & 13.2 &  6.9 &  24&          & $ W2 $ & 35 & 32.4 &  10.6 & 134 \\
         &$S^\ast$&  88 &      &      &    & 1242+410  & $ S $ & 818 &20.8 &  5.1 & 26 \\             
0147+400 & $ E  $ & 208 &  2.7 &  2.4 & 41 &           & $ N $ & 208 &7.7 &  4.8 & 53 \\              
         & $ W1 $ & 134 & 11.6 &  7.6 & 40 &           &$T^\ast$ & 101 &   &     &    \\              
         & $ W2 $ & 109 & 13.7 & 10.6 & 26 & 1314+453  & $ W1 $ & 175 & 26.7 & 15.6 &  71 \\          
 	 &$T^\ast$& 113 &      &      &    &           & $ W2 $ & 77 &  31.6 & 26.9 &  73 \\          
0703+468 & $ W1 $ & 794 &  7.1 &  4.01& 51 &           & $ Ce $ & 60 &  19.0 & 15.0 & 143 \\          
	 &$W^\ast$&  22 &      &      &    &           & $ E1 $ & 101 & 28.0 & 14.6 &  63 \\          
         & $ E1 $ & 497 &  5.7 &  4.0 & 80 &           & $ E2 $ & 76 &  41.4 & 10.2 &  65 \\          
	 &$E^\ast$&  44 &      &      &    &           & $ NW  $ & 18 &  25.0 &  8.9 &  82 \\         
0800+472 & $ S1 $ & 177 & 25.5 &  9.4 &  49& 1340+439  & $ N1 $ & 208 & 9.0&  5.2 &   9 \\            
         & $ S2 $ &  95 & 32.4 &  6.9 &  70&           & $ N2 $ & 104 &  13.9 &  5.8 &  44 \\         
         & $ N1 $ & 120 & 36.5 & 19.2 &  34&           & $ Ce1 $ & 38 &  10.3 &  1.5 &  26 \\         
         & $ N2 $ &  68 & 25.6 & 22.9 &  60&           & $ Ce2 $ & 13 &  7.2 &  2.2 &   4 \\          
         & $ N3 $ &  24 & 18.8 & 11.3 & 178&           & $ S  $ & 27 &  12.0 &  6.6 &  83 \\          
0809+404 & $ N $  & 250 & 16.1 & 12.2 & 161& 1343+386  & $ S1 $ & 541 & 6.2 &   5.9 & 131 \\          
         & $ S $  & 221 & 17.5 & 12.2 & 161&           &$S^\ast$ & 68 &     &       &     \\          
         &$T^\ast$&  51 &      &      &    &           & $ Ce  $ & 20 &  12.8 &   4.5 & 170 \\        
0822+394 & $ E1 $ & 424 &  5.3 &  3.5 &  37&           & $ N  $ & 66 &  3.5 &   0.6 & 171 \\          
         & $ E2 $ & 269 & 16.4 &  3.3 &  57&           &$N^\ast$ & 20 &     &       &   \\            
         & $ W1 $ & 153 &  6.7 &  4.5 &  66& 1432+428  & $ E1 $ & 570 & 4.9 &  3.0 & 106 \\           
         & $ W2 $ &  97 &  6.4 &  3.7 &  81&           &$E^\ast$ &  25 &    &      &     \\           
0840+424 & $ N1 $ & 823 &  8.2 &  4.8 &  10&           & $ W  $ & 104 & 11.4 & 7.5 & 133 \\           
	 & $ N2 $ & 145 & 12.5 &  4.5 & 169& 1441+409  & $ W1 $ & 312 & 6.7 &  4.2 &   50 \\          
         & $ S1 $ &  71 &  7.1 &  5.7 &  50&           & $ W2 $ & 132 & 17.1 &  8.0 &   43 \\         
         & $ S2 $ & 137 & 16.8 & 11.6 & 139&           & $ Ce1 $ & 140 & 7.6 &  3.0 &   63 \\         
1007+422 & $ S1 $ & 238 & 32.9 & 15.8 &  84&           & $ Ce2 $ & 42 &   $<3.5$ & $<3.5$ & -180 \\   
         & $ S2 $ &  52 & 22.4 &  8.7 &  40&           & $ E1 $ & 70 & 8.0 &  3.1 &  176 \\           
         & $ N  $ &  32 & 40.0 & 23.5 & 153&           & $ E2 $ & 68 & 17.2 &  7.3 &   30 \\          
1008+423 & $ W1 $ & 139 & 12.3 &  4.4 & 128& 1449+421 & $ N1 $ & 121 & 12.04 &  7.0 & 96 \\           
         & $ W2 $ & 250 & 12.0 &  5.3 &  35&          & $ N2 $ & 149 & 5.2 &  2.7 & 26 \\             
	 &$W^\ast$&  14 &      &      &    &          & $ N3 $ & 87 &  8.09 &  4.1 &  4 \\            
         & $ Ce $ &  14 & 12.2 &  5.2 &  54&          & $ Ce1 $ & 15 &  8.5 &  2.6 & 24 \\            
         & $ E $  & 104 & 27.8 & 19.4 &  46&          & $ Ce2 $ & 58 &  5.8 &  3.4 & 31 \\            
1016+443 & $ N1 $ & 109 &  8.1 &  5.5 & 156&          & $ Ce3 $ & 44 &  8.9 & 2.1 & 27 \\             
         & $ N2 $ &  67 & 15.1 &  3.5 & 175&          & $ S  $ & 120 & 11.0 & 8.2 & 60 \\             
         & $ N3 $ &   6 & 15.9 &  1.6 & 166& 2304+377 & $ W1 $ & 244 & 4.8 &  0.2 &  89 \\            
         & $ S1 $ &  27 & 15.1 &  7.7 & 155&          & $ W2 $ & 671 & 21.0 &  6.6 & 117 \\           
         & $ S2 $ &  58 & 18.9 &  7.2 &   5&          & $ W3 $ & 91 &  7.1 &  2.3 & 104 \\            
1044+454 & $ E1 $ & 124 &  8.3 &  3.8 & 106&          & $ E1 $ & 219 &  4.0 &  3.3 & 114 \\           
         & $ E2 $ & 138 & 19.9 & 17.1 & 110&          & $ E2 $ & 41 &  9.0 &  5.6 &  67 \\            
	 &$E^\ast$&  18 &      &      &    &          & $ E3 $ & 124 & 28.0 & 17.8 & 110 \\           
         & $ W1 $ &  24 &  7.3 &  4.4 &   6& 2330+402 & $ W1 $ & 332 & 2.4 &  1.8 &  33 \\            
         & $ W2 $ &   7 & 10.6 &  1.7 & 177&          &$W^\ast$ & 37 &     &      &     \\            
1049+384 & $ W1 $ & 244 &  7.0 &  4.1 &   8&          & $ E1 $ & 62 & 7.0 &  4.8 & 152 \\             
         & $ W2 $ & 206 &  8.7 &  2.5 & 119&          & $ E2 $ &204 & 17.5 & 12.9 & 164 \\            
         & $ E1 $ &  21 &$<5.5$&$<5.5$&-179&          & $E^\ast$ &70 &     &      &     \\            
         & $ E2 $ &  26 &  5.4 &  3.6 & 133& 2348+450 & $ S1 $ & 203 &5.1 &  4.1 & 159 \\             
1133+432 & $ N $  & 799 &  3.5 &  2.9 &  48&          &$S^\ast$ & 168 &   &      &     \\             
         & $ S $  & 460 &  4.1 &  3.6 & 167&          & $ N1 $ & 75 & 19.9 & 14.4 &  93 \\            
1136+383 & $ N1 $ & 134 &  5.2 &  3.0 & 170&          & $N^\ast$ &74 &     &      &     \\            
         & $ N2 $ &  93 & 10.1 &  4.1 &  10& 2358+406 & $ S1 $ & 681 &5.6 &  3.9 & 143 \\             
         & $ Ce $ &  23 &  8.2 &  3.2 & 170&          &$S^\ast$ & 158 &   &      &     \\             
         & $ S  $ & 125 &  5.8 &  5.1 &   3&          & $ Ce  $ & 71 & 9.3 &  1.0 & 156 \\            
1136+420 & T      & 292 & 52.2 & 31.1 & 103&          & $ N1 $ & 95 & 8.8 &  4.3 &   3 \\             
1159+395 & $ N1 $ & 198 &  5.2 &  3.3 & 179&          & $ N2 $ & 213 &15.1 &  6.1 & 123 \\            
         & $ N2 $ & 130 & 21.3 &  9.4 & 167&          &        &     &&    &   \\
         & $ S1 $ & 165 &  6.7 &  4.8 & 141&          &        &     &&    &   \\
\hline
\hline
\end{tabular}
\label{comparam}
\end{center}
\end{table*}

\section{Comments on individual sources}
\label{comm}
-- 0800+472: the source has an overall extent of~$\sim$ 1 arcsec in the VLA
8.4 GHz image (see paper I).  About 40 \% of the flux density is
missing in our VLBA image. Since $\sim$ 90 \% of the 8.4 GHz flux density
originates within a region of $\sim$ 120 mas, it is likely that most of the 
missing flux density is on angular scales of this order.
The VLBA source structure is strongly bent and the bending probably
continues to match the arcsec structure. It is possible that we are in presence
of an asymmetric core--jet, with a weak core, as in Stanghellini et al.
(\cite{stan}).

\smallskip
\noindent
-- 0809+404: at arcsec scale the source appears as a very
asymmetric double, with 1.2$^{\prime\prime}$ 
separation and a flux density ratio $\sim$ 100:1 at 4.9 GHz (see paper I).
The structure we see here corresponds to the brighter component and is
rather amorphous, with the p.a. of the major axis misaligned by $\sim$
40$^\circ$ with respect to the orientation of the outer weak
component. About 45 \% of the total flux density is missing in our
VLBA observations.  

\smallskip
\noindent
-- 1007+422: about 15 \% of total flux density is missing in our VLBA 
observations. 

\smallskip
\noindent
-- 1044+454: on the arcsec scale the source appears as a very asymmetric
double, with flux density ratio $\sim$ 20:1 at 4.9 GHz (see paper I).
On the VLBA scale the structure of the northernmost brighter component is
again double and asymmetric, along the same p.a. but on a $\sim$ six times
smaller scale. The source could be either a core-jet with a weak core or
a very asymmetric MSO.
About 12 \% of the total flux density is missing in 
our VLBA observations 

\smallskip
\noindent
-- 1049+384: About 12 \% of the total flux density is missing in 
our VLBA observations 

\smallskip
\noindent
-- 1136+420: on the arcsec scale the source appears as a very asymmetric 
double, with flux density ratio $\sim$ 20:1 at 4.9 GHz (see paper I),
roughly oriented in the East--West direction. At VLBA resolution only the brighter component
is visible and displays a very amorphous structure.
About 30 \% of the total flux density is missing in 
our VLBA observations 

\smallskip
\noindent
-- 1314+453: about 15 \% of the total flux density is missing in the VLBA 
image.

\smallskip
\noindent
-- 1343+386: a weak feature, accounting for $\sim$ 9 mJy is present to
the North of
the major components.

\begin{figure}
\vspace{9.5 truecm}
\includegraphics{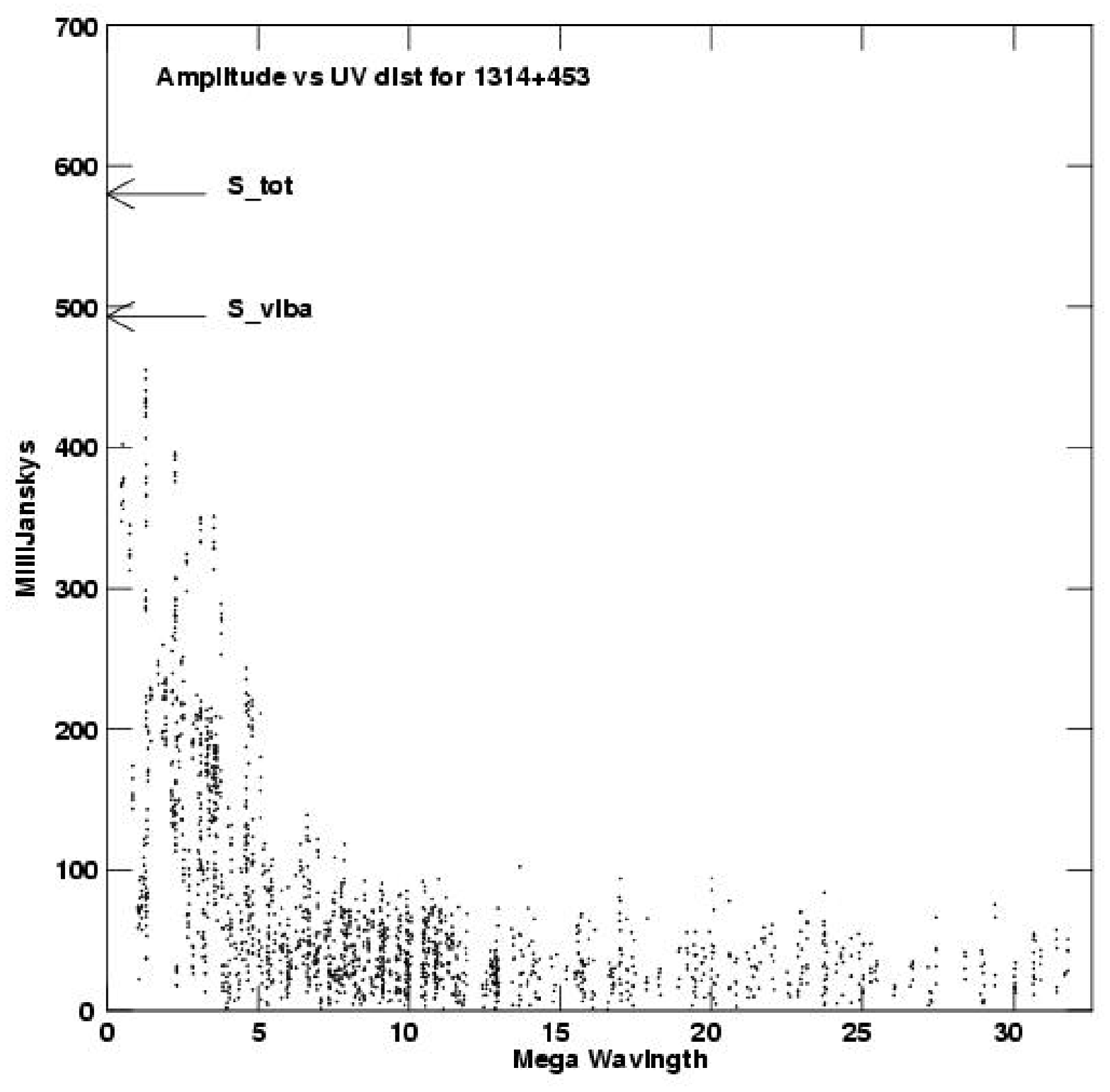}
\caption{Plot of the correlated flux density versus projected
baseline for the source B3 1314$+$453. The horizontal arrows
indicate the integrated flux density in the VLBA image and total flux
density from the VLA as given in Table 1}
\label{fluxbas}
\end{figure}

\smallskip
\noindent
-- 1432+428: a weak feature, accounting for $\sim$ 10 mJy is present
to the South of the major components. About 13 \% of the total flux
density is missing in our VLBA observations 

\smallskip
\noindent
-- 2348+450: about 20 \% of total flux density is missing in the VLBA 
image.

\section{Discussion}
\label{morph}
The VLBA observations have produced images with sufficient resolution
to describe the source morphology for the large majority of cases.

The present observations do not allow to pick up the radio core, due to the
relatively low observing frequency, but the overall morphology strongly
indicates that the sources are mostly two--sided. We see either well
separated lobes or lobes connected by bridges, possibly jets. 
Most sources are edge brightened, suggesting the presence of hot spots,
in general not fully  resolved from the mini--lobe emission, which 
contribute most of the source flux density.  We tentatively classify
these sources as CSOs (two of them are actually MSOs).
There are also a few objects with edge darkening, as B3 2304+377, B3 2348+450
and B3 2358+406.
There are a number of sources where a compact component, centrally
located, could be the source core. However, having only one
relatively low frequency we are not in the position to make a
definitive statement about core identifications. 

Some asymmetries in the flux densities of the two sides are seen, as
also some moderate distortions. B3 0800+472 and B3 2304+377 are
extreme examples. The sources B3~1225+442 and B3 1449+421 are
reminiscent of an $S$ morphology (see also Stanghellini et
al. \cite{stan}). 

The average  ratio between the component size transverse to the source
axis ($\theta_{\rm trans}$) and Largest Angular Size ($LAS$) is
$\theta_{\rm trans}/LAS \sim 0.08 \pm 0.02$, consistent with findings
in other source samples (e.g. Fanti et al. \cite {fan90})

We have computed the equipartition parameters for the source
components, under the following assumptions: a) proton to electron energy
ratio of one; b) filling factor of one; c) maximum and minimum
electron (and proton) energies  corresponding to synchrotron emission
in the frequency range 100 GHz to 10 MHz; d) ellipsoidal volumes with
axis corresponding to the observed ones. We also computed the
brightness temperatures ($T_{\rm B}$) of each component.  
In order to make easier the reading of the paper, we do not report the
individual component values, since they can be obtained from the
parameters in  Table \ref{comparam}, but only mention the typical
values. 

The magnetic fields are in the range of several mGauss, going up to
$\sim$ 20 mGauss in the most compact components
(e.g. B3~1133+432). Brightness temperatures range from $\sim 10^8$ up
to $\leq 10^{11}$ K. 

\section{Conclusions and future plans} 
We have presented the results of VLBA observations of the most compact
sources ($LAS \ltsim 1\arcsec $) in the B3--VLBA CSS sample from paper I.

The large majority of sources have a ``double type" structure, with
sizes ranging from $\sim$ 200 $h^{-1}$ pc to $\sim$  1 $h^{-1}$ kpc.  
Together with the VLA data on the large size sources and with EVN \&
MERLIN data, presented in Dallacasa et al. (\cite{dalla}, Paper III)
on intermediate  size sources, we have at the end a new determination
of the Linear Size distribution in the range $\sim 0.2~h^{-1} <
LS$(kpc) $<$ 20 $h^{-1}$. This is an important piece of information
to test the source evolution model. We will discuss it in a separate
paper. 

The sources presented here have just been observed with the VLBA at
higher frequencies, with the aim of detecting the radio core,
therefore obtaining an unambiguous classification, and of studying the
structure of the hot spots. The results will be presented in a
forthcoming paper.

\begin{acknowledgements} 
This work has been partially supported by the Italian MURST under
grant COFIN-2001-02-8773.
The VLBA is operated by the U.S. National Radio Astronomy Observatory 
which is a facility of the National Science Foundation operated under
a cooperative agreement by Associated Universities, Inc.
\end{acknowledgements}

\end{document}